\documentclass[12pt]{iopart}
\usepackage{graphicx}
\usepackage{iopams}  

\expandafter\let\csname equation*\endcsname\relax
\expandafter\let\csname endequation*\endcsname\relax
\usepackage{amsmath}
\usepackage{cite}
\begin{document}

\title{photon echo with a few photons in two-level atoms }

\author{M. Bonarota, J. Dajczgewand, A. Louchet-Chauvet,\\ J.-L. Le Gou\"et, T. Chaneli\`ere }

\address{Laboratoire Aim\'e Cotton, CNRS-UPR 3321, Univ. Paris-Sud, B\^at. 505, 91405 Orsay cedex, France}
\ead{jean-louis.legouet@lac.u-psud.fr}
\begin{abstract}
To store and retrieve signals at the single photon level, various photon echo schemes have resorted to complex preparation steps involving ancillary shelving states in multi-level atoms. For the first time, we experimentally demonstrate photon echo operation at such a low signal intensity without any preparation step, which allows us to work with mere two-level atoms. This simplified approach relies on the so-coined "Revival Of Silenced Echo" (ROSE) scheme. Low noise conditions are obtained by returning the atoms to the ground state before the echo emission. In the present paper we manage ROSE in photon counting conditions, showing that very strong control fields can be compatible with extremely weak signals, making ROSE consistent with quantum memory requirements.    
\end{abstract}
\maketitle

\section{Introduction}
For the last two decades, the quantum information prospects have stimulated the investigation of atom-based quantum memory (QM) for light. The QM challenge includes the conversion of a quantum state of light into an ensemble state of matter and the retrieval of a restored state of light. Light operates as a probe of the ensemble entangled state. Electromagnetically induced transparency (EIT) combined with the DLCZ generation of narrow-band heralded photons led to convincing experimental demonstrations in atomic vapors~\cite{Chaneliere2005,Eisaman2005,appel2008}. 

Boasting a rich background in classical signal processing, photon echo (PE) was also recognized as a rather obvious QM candidate, enjoying attractive bandwidth and multimode capabilities. To objections regarding noise, recovery efficiency and fidelity, a seminal paper replied, showing the promises of a specific PE scheme~\cite{Moiseev2001}, and so demonstrating the absence of fundamental limitations in PE application to quantum storage of light. This really triggered an intense effort of investigation, leading to the emergence of different PE schemes~\cite{nilsson2005,Hetet2008bis,afzelius2009} and to PE-based memory demonstrations at the single photon level~\cite{Riedmatten2008,Chaneliere2010,sabooni2010,Lauritzen2010,Timoney2013,guandogan2013}, then with quantum light such as heralded single photons~\cite{saglamyurek2011,clausen2011}. 

Meanwhile, weak classical signal PE-based storage was being studied~\cite{Tsang2003} and the limitations of two-pulse photon echo (2PE) for quantum storage were clarified~\cite{ruggiero2009,sangouard2010,ledingham2010}. The intense light pulse that reverses the phase of atomic coherences and get them phased together at a later time, simultaneously promotes the atoms to the upper level of the optical transition. Working in a gain regime, detrimental to fidelity, the inverted medium also relaxes by spontaneous emission (SE), which further increases the intrinsic noise and makes the scheme improper to the recovery of the initial quantum state of light. Last but not least, it was realized that the recovered signal was contaminated by a free induction decay (FID) emission. Indeed one cannot avoid the generation of a large residual ensemble coherence by the strong rephasing pulse. The resulting undesired coherent radiation is emitted in the same spatial and frequency mode as the rephasing pulse~\cite{Beavan2011}, quite close to the echo direction, according to the 2PE phase matching conditions.     
To adapt the photon echo to quantum memory requirements, one has to get rid of massive population inversion. All the successful PE-based schemes, including the controlled reversible inhomogeneous broadening (CRIB)~\cite{moiseev2004,nilsson2005,kraus2006,sangouard2007}, the gradient echo memory (GEM)~\cite{alexander2006,Hetet2008,Hetet2008bis,Longdell2008,hedges2010,Lauritzen2010} and the atomic frequency comb (AFC)~\cite{afzelius2009,Riedmatten2008,Chaneliere2010,sabooni2010}, dispose of spontaneous emission (SE) noise by avoiding massive population inversion. To reach this goal, they all resort to ancillary states, where spectrally selected atoms are shelved before memory operation. In addition to complicating the storage scheme, this preparation step reduces the intrinsic capture capability of the medium by removing atoms from the absorption band and strongly reducing the available optical depth. 

Among the rich harvest of AFC results, most of them have been obtained in a simplified scheme where the storage medium operates as a pre-programmed delay line, rather than as an on-demand memory. The single photon level have been reached quite recently in on-demand memory conditions~\cite{Timoney2013,guandogan2013}, where strong pulses are needed. Managing intense light in a single-photon detection context remains problematic, although those strong pulses only excite transitions between nearly empty levels. 

Returning to 2PE basics, we proposed an alternative approach, free from any preparation step, with on-demand capacity, and going without ancillary states~\cite{damon2011}. After a first strong rephasing pulse, that excites the atoms to the upper level, a second intense pulse brings the atoms back to the ground state, thus suppressing the SE noise. Simultaneously, the second strong pulse reverses the phase of the atoms, making them emit a secondary echo. The beam directions play a key role in this memory-oriented photon echo scheme. The two strong pulses propagate colinearly, in the opposite direction to the signal to be stored. This beam arrangement is spatially phase mismatched for the primary 2PE emission, but fully phase matched for the secondary echo emission. Since the primary 2PE is silenced, the integrity of the captured information is preserved until the emission of the secondary echo. At the same time the counterpropagating beam arrangement eliminates the FID tail that travels along with the rephasing pulses. 
 
The so-coined "Revival Of Silenced Echo" (ROSE) scheme clearly disposes of two insurmountable faults of the conventional 2PE, namely the spontaneous emission (SE) noise and the signal contamination by the FID tail of the rephasing pulses. Unlike the protocols based on a preparation step, such as CRIB, GEM or AFC, the ROSE does not waste the available optical thickness. All the atoms initially present within the signal bandwidth may participate to the quantum memory. Finally, since the shelving states are not needed, the ROSE can work in two-level atoms, when long coherence lifetime can be sacrificed for large multimode capacity. 

Taking advantage of those properties, we presently demonstrate photon echo operation with a few photons in a two-level system, for the first time ever to the best of our knowledge. After a brief summary of ROSE basics in Section~\ref{section:protocol}, we describe the experimental setup in Section~\ref{section:setup}. The noise features are examined in Section~\ref{section:noise}, and the ROSE operation with 15 photons is presented in Section~\ref{section:low_level_ROSE}.    
   
\section{Revival of silenced echo (ROSE)}\label{section:protocol}
\begin{figure}[h!]
\centering{\includegraphics[width=12cm]{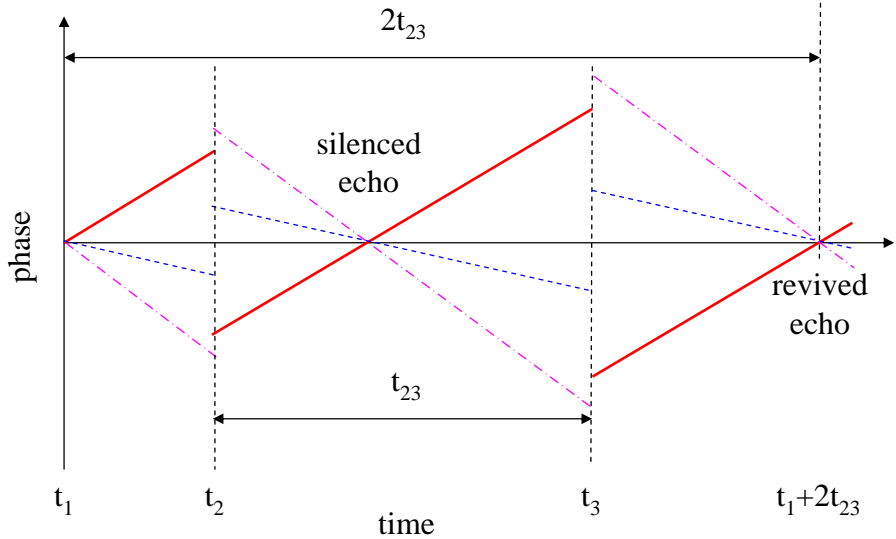}}
\caption{Revival Of Silenced Echo. Excitation at time $t_1$ gives rise to atomic coherences. Departing from their initial common phase, the atomic coherences evolve at different rates, depending on their detuning from a reference. Rephasing pulses are shone at times $t_2$ and $t_3$. The atomic coherences get phased together at time $t_1+2t_{12}$ but the primary echo is silenced by spatial phase mismatching. The echo is revived at time $t_1+2t_{23}$.}
\label{fig:ROSE}
\end{figure}

The ROSE protocol has been detailed in Ref.\cite{damon2011}. Let us briefly summarize a few key features. The ROSE scheme directly derives from the two-pulse echo (2PE). As such, ROSE operates on an inhomogeneously broadened optical transition, involving an ensemble of two-level atoms. A weak pulse, carrying the information to be stored, shines the atomic medium at time $t_1$ and, provided the optical thickness is large enough, is fully converted into atomic coherences. Evolving at different frequencies, the coherences build up different spectral phase shifts. Then a strong pulse hits the medium at time $t_2$, simultaneously reversing the spectral phase shift of the coherences and promoting the atoms to the upper level. The different spectral phase shifts corresponding to coherences at different transition frequencies simultaneously vanish at time $t_e=t_1+2t_{12}$, where $t_{ij}=t_j-t_i$. In conventional 2PE, in-phase coherences radiate an echo at time $t_e$ but, since the medium is fully inverted, the echo is contamimated by fluorescence. In ROSE, spatial mismatching is used to prevent echo emission and to keep information safe inside the medium, until a second strong pulse, applied at time $t_3>t_e$, brings the atoms back to the ground state and, phasing the coherences together again, let an echo be radiated at time $t_{e'}=t_1+2t_{23}$ from ground state atoms, free of fluorescence noise (see Fig.~\ref{fig:ROSE}).  

The atomic coherences simultaneously carry a spatial and a spectral phase, both of them being controlled by the driving pulses. Hence the coherences are spatially distributed as a matter wave. To be emitted by the atomic coherences, the light wave signal must be phase matched to the matter wave. In conventional 2PE, the pulses, directed along wavevectors $\vec{k}_1$ and $\vec{k}_2$, convey the spatial phase $\vec{k}_{e}\cdot\vec{r}$ to the coherences, where $\vec{k}_e=2\vec{k}_2-\vec{k}_1$. An echo is emitted in direction $\vec{k}_e$, provided $k_e=|\vec{k}_e|$ is close to $k=|\vec{k}_1|=|\vec{k}_2|$. More precisely, the phase matching condition reads: $|k_e-k|L<<\pi$, where $L$ stands for the medium thickness. In ROSE, the first two pulse wavevectors are directed in such a way that $|k_e-k|L>\pi$. Hence the coherences keep silent, no echo being emitted at time $t_e$. The strong pulse at time $t_3$ reverses the spectral phase shift, making all the excited coherences recover a common spectral phase at time $t_{e'}=t_1+2t_{23}$, and simultaneously converts the spatial phase $\vec{k}_{e}\cdot\vec{r}$ into $\vec{k}_{e'}\cdot\vec{r}$, where $\vec{k}_{e'}=2\vec{k}_3-\vec{k}_e=\vec{k}_1+2\left(\vec{k}_3-\vec{k}_2\right)$. A secondary echo is emitted at time $t_{e'}$ provided $|k_{e'}-k|L<<\pi$. If $\vec{k}_3=\vec{k}_2$, $\vec{k}_{e'}$ reduces to $\vec{k}_1$ and $|k_{e'}-k|L=0$. Hence the phase matching condition, violated at time $t_{e}$, is fully satisfied at time $t_{e'}$ and emission takes place in the same $\vec{k}_1$ direction as the initial signal, whatever the common wavevector direction of the two rephasing pulses. This occurs for instance when the two rephasing pulses counterpropagate with the incoming signal. The 2PE emission is strictly forbidden since $|k_e-k|=2k$ and the different beams overlap exactly, irrespective of the sample length. 

\begin{figure}[h!]
\centering{\includegraphics[width=10cm]{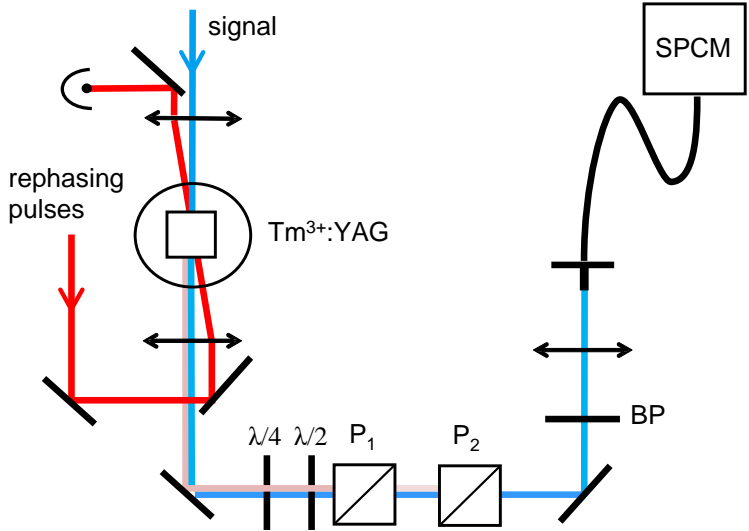}}
\caption{ Setup. The signal field is cross-polarized with respect to the rephasing pulses. The rejection of stray light from the latter beams by polarizers $P_1$ and $P_2$ is optimized by the waveplates $\lambda/4$ and $\lambda/2$. The polarizer pair offers a nominal extinction ratio of 10$^{-7}$. The echo is detected on a single photon counting module (Perkin Elmer SPCM-AQR-14). Spatial filtering through an optical fiber isolates the relevant mode. In addition, spectral filtering by a bandpass (BP) filter only keeps the transition to the lower sublevel of the ground state Stark multiplet. }
\label{fig:setup}
\end{figure}
 
As they counterpropagate with the signal, the rephasing pulses no longer contaminate the echo with their FID tail, as they do in conventional 2PE~\cite{ruggiero2009}. More problematic is the residual SE, when all the atoms are not returned to the ground state by the second intense pulse. In conventional 2PE, when the medium is totally inverted by the rephasing pulse, SE brings a noise of about 1 photon within the temporal mode and the solid angle of echo emission~\cite{ruggiero2009}. More generally, the SE photon rate in the echo mode can be expressed as:
\begin{equation}\label{SE_rate} 
R_{SE}=\alpha L n_e \Delta/\pi
\end{equation}    
where $\alpha L$, $n_e$ and $\Delta$ respectively stand for the optical depth, the upper level population and the spectral width of the inverted region, expressed in angular frequency units. This SE rate expression is valid if $\alpha L\leq1$. Extension to larger optical depth values can be found in Ref.~\cite{ledingham2010}. As expressed in terms of the absorption coefficient from the ground state, the rate given by Eq.~(\ref{SE_rate}) only reflects direct decay to the ground state. Decay channels to other states may bring additional contributions to SE background if they are not filtered out properly. In a similar way, Eq.~(\ref{SE_rate}) refers to a single polarization state. The observed SE rate is higher in the absence of polarizer before the detector. 
  
\section{Experimental setup}\label{section:setup}
The experiment is conducted in a 8 mm-long, $0.1\%$ at. Tm$^{3+}$:YAG crystal operating at 793nm. The concentration is lower than in Ref.~\cite{damon2011}, which should lead to longer coherence lifetime. The sample is cooled down to 2.6K in a variable temperature liquid helium cryostat. At the line center, $\alpha L$ is close to 1.4. The counterpropagating light beams are split from a home-made, extended cavity, continuous wave, ultra-stable semi-conductor laser~\cite{crozatier2004}, featuring a stability better than 1 kHz over 10 milliseconds.   

\begin{figure}[h!]
\centering{\includegraphics[width=10cm]{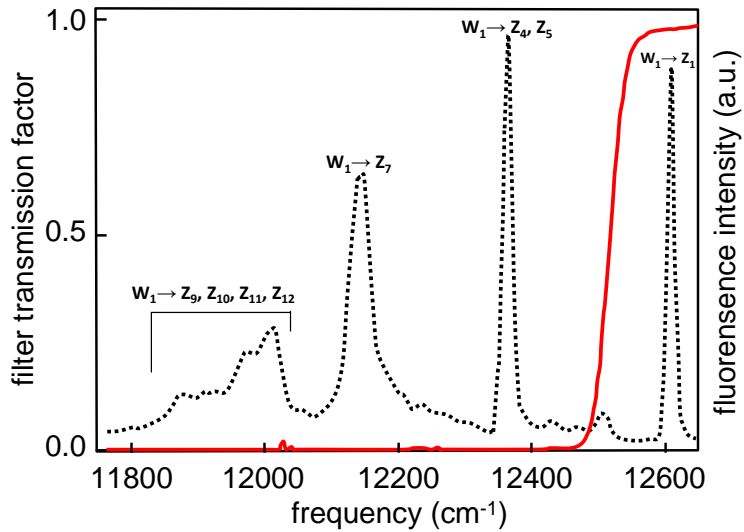}}
\caption{Spectral filtering of fluorescence. Out of the fluorescence spectrum of the $^3$H$_4\rightarrow^3$H$_6$ transition (dotted line)~\cite{Gorju2007}, the Semrock bandpass filter (red solid line) selects the highest frequency line, connecting the lower levels of the Stark multiplets in upper and ground states. The represented fluorescence lines connect $W_1=^3$H$_4(0)$, the lower Stark level of the upper state, to $Z_{n+1}=^3$H$_6(n)$, the different Stark sublevels of the ground state. }
\label{fig:RE_filtering}
\end{figure}

As illustrated in Fig.~(\ref{fig:setup}), various techniques have been involved in the elimination of spurious light. The rephasing pulses counterpropagate with the signal, in accordance with the ROSE phase-matching conditions. A small angle between the two paths allows us to extract the echo easily. By cross-polarizing the incoming signal and the echo with the rephasing beams, we further reject the intense pulse reflection off the cryostat windows and the crystal surface. The rephasing pulses and the signal are respectively polarized along crystallographic axes $[\bar{1}10]$ and $[001]$~\cite{sun2000}. Thulium ions are substituted to yttrium in 6 orientationnaly inequivalent sites. With the selected polarization arrangement, four sites only contribute to the input pulse capture. Those four sites behave equivalently with respect to both the signal and the rephasing pulses~\cite{bonarota2011}. The signal beam waist is adjusted to 30$\mu$m, with a rephasing beam 1.6 times bigger. Spatially filtered by a single mode optical fiber, only radiation in the incoming signal spatial mode can reach the single photon counting module (SPCM). Finally, as shown on Fig.~(\ref{fig:RE_filtering}), a band-pass filter (Semrock FF01-786/22) selects the $^3$H$_4(0)\rightarrow^3$H$_6(0)$ transition connecting the lower levels of the Stark multiplets, allowing the rejection of fluorescence from $^3$H$_4(0)$ to $^3$H$_6(n\neq0)$ sublevels. Experimentally, the filter reduces fluorescence intensity at the detector by factor $\approx7.5$.   

Phase reversal is achieved by Complex Hyperbolic Secant (CHS) pulses, a variant of Adiabatic Rapid Passage~\cite{deSeze2005,rippe2005}. The time variation of the CHS Rabi frequency is given by:
\begin{equation}
\Omega(t)=\Omega_0\{\text{sech}\left(\beta(t-t_0)\right)\}^{1-i\mu}
\end{equation}   
Such a pulse is able to flip the Bloch vector over a $2\mu\beta$-wide spectral interval provided $\mu>2$ and $\mu\beta^2<<\Omega_0^2$~\cite{deSeze2005}. Defined in a less restrictive way than $\pi$-pulses, the CHS pulse is also less sensitive to propagation-induced distorsion~\cite{Spano1988} and has proved efficient in the ROSE context~\cite{damon2011}. The rather short $T_2$ ($<$ 100$\mu$s) impacts on the accessible parameter range, while $\Omega_0$ is limited by the available laser power. To simultaneously satisfy $T_2\beta>>1$ and  $\mu\beta^2<<\Omega_0^2$ (adiabatic passage condition), one is led to $\mu=3$, with $\beta=2\pi\times80$ kHz. Hence level population is inverted over a $2\mu\beta=2\pi\times480$ kHz-wide spectral interval.  

\section{Spontaneous emission and coherent noise}\label{section:noise}
First we observe SE from the fully inverted medium, after excitation by a single CHS pulse. The medium response temporal evolution is explored by counting photons from the SPCM with a time digitizer (P7888 FAST ComTec). The time window is 256-ns-wide, giving a resolution shorter than the signal duration. Each curve is obtained by integrating 15000 ROSE sequences corresponding to a 300s experimental run (50Hz repetition rate). 
\begin{figure}[h!]
\centering{\includegraphics[width=10cm]{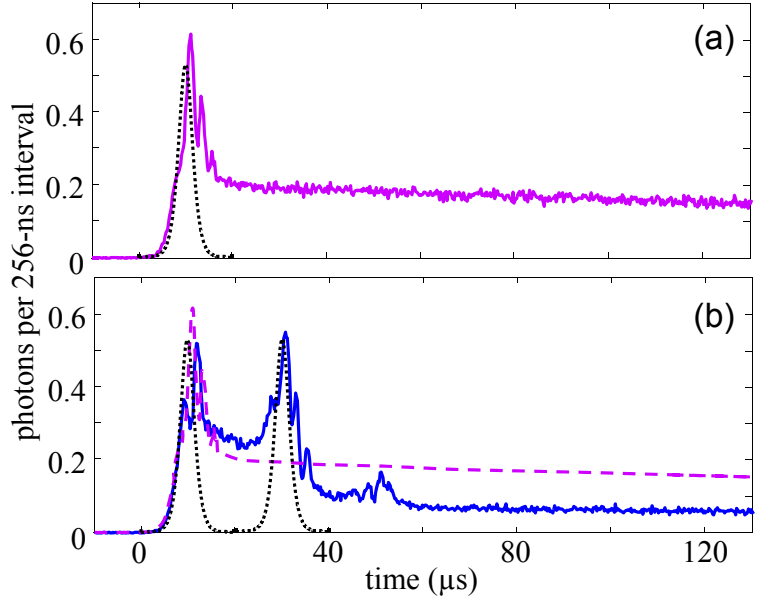}}
\caption{Medium response after excitation by one (a) and two (b) CHS pulses, with $\alpha L\approx1.4$. The detected counts in a 256-ns time window are represented per ROSE sequence in the crystal (i.e. normalized by the overall detection efficiency, 55\% $\times$ 40\%, see text for details). Error bars can be estimated from the statistics of photo-counting events: as a reference, 0.2 photon corresponds to 646 detected events. The dashed line in (b) reproduces the curve displayed in (a). The dotted lines represent the temporal position and shape of the CHS pulses. They are scaled to the height of the observed stray light.}
\label{fig:inversion_noise}
\end{figure}

In order to gain access to the number of counts in the crystal, we normalize by the overall detection efficiency. The latter is composed of the SPCM photodetection efficiency ($\sim$ 55\%) and the collection effiency from the crystal to the dectector ($\sim$ 40\% essentially limited by the uncoated faces of the crystal, the cryostat viewports and the fiber coupling to the SPCM). To summarize, the total recorded counts divided by 15000 $\times$ 55\% $\times$ 40\%, corresponds to the number of counts per ROSE sequence in the crystal. As shown in Fig.~(\ref{fig:inversion_noise}.a), some stray light from the CHS pulse is followed by a long tail of spontaneous emission, decaying in an exponential way with characteristic time $T_1=460\pm60\mu$s, consistent with the upper level lifetime. At maximum, one detects $\approx0.2$ SE photons per 256-ns interval, a little less than predicted by Eq.~(\ref{SE_rate}).    

The response after excitation by two CHS pulses is represented in Fig.~(\ref{fig:inversion_noise}.b). After the second CHS pulse the SE tail is about 3 times smaller than in single CHS pulse conditions. Hence $\approx$2/3 of the ions have been returned to the ground state, but $\approx$1/3 are still left in the upper level. An additional emission can be observed in the wing of the second pulse. This may be identified either as the backward reflection of a photon echo produced by the CHS pulses, or as the echo produced by the backward reflected CHS pulses. Anti-reflection coating the crystal would reduce this coherent emission. 

At this stage we must admit we are not yet able to offer the ROSE a vanishing background. However significant progress have been accomplished toward this goal. A few microseconds away from the strong rephasing pulses, each carrying about 10$^{12}$ photons, and flipping nearly 10$^{11}$ Tm ions from the ground state to the upper level and back, we have reduced the background to about 1 photon per echo mode. This level is low enough for a ROSE demonstration with a few photon signal.             

\section{Low level ROSE}\label{section:low_level_ROSE}
\begin{figure}[h!]
\centering{\includegraphics[width=10cm]{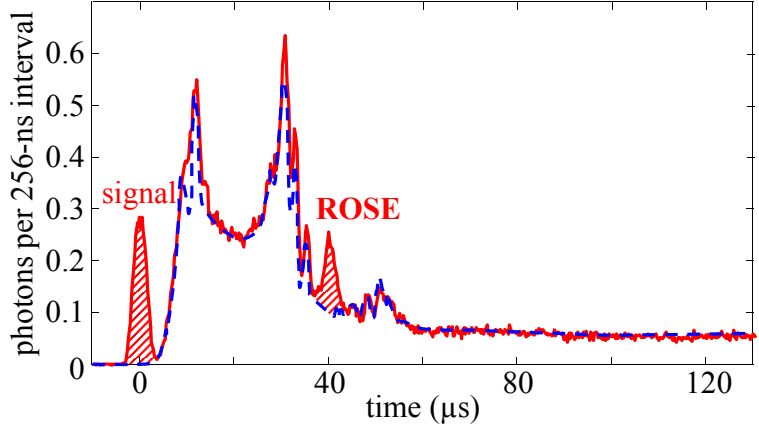}}
\caption{ROSE with 14 photons. The time evolution is explored step by step through the 256-ns-wide window of the SPCM. The dashed line reproduces the two-CHS excitation curve without incoming signal, as displayed in Fig.~(\ref{fig:inversion_noise}.b). The shaded areas represent the signal, incoming at $t=0$, and the ROSE, recovered at $t=40\mu$s. The experimental conditions are the same as in Fig.~(\ref{fig:inversion_noise}).}
\label{fig:low_level_ROSE}
\end{figure}

The incoming data pulse intensity is given the gaussian shape $\exp(-t^2/\tau^2)$, with $\tau=2\mu$s. It contains 14 photons. This number is estimated by integrating the counts in fig. 5 within the gaussian mode of the transmitted signal. It corresponds to 3.5 photons that a not captured by the medium due to  the limited optical depth. After dividing by $\exp(-\alpha L)$, we obtain the incoming photon number of 14. To calculate the signal-to-noise ratio at the retrieval stage, we apply the same procedure. The gaussian integration over the ROSE temporal profile ( fig. 5 ) gives 1.4 photons in the echo. As a comparison, the noise background in dashed line after a gaussian averaging gives 1.1 photons. In those conditions, the signal-to-background ratio at the retrieval is close to unity. According to Fig.(~\ref{fig:low_level_ROSE}), where the incident signal and the ROSE are represented by shaded areas, the recovery efficiency reaches $\approx12\%$. This is consistent with the $(\alpha L)^2\rme^{-\alpha L-4t_{23}/T_2}$ expected value, with $t_{23}=20 \mu$s and $T_2=55 \mu$s. 

With $\tau=2\mu$s, the incoming signal spectrum fits within the 480 kHz bandwidth of the CHS rephasing pulses. Hence the recovered shape satisfactorily coincides with the incident one. If some shape distortion is acceptable, one can improve the signal-to-background ratio by using a shorter input pulse. For instance a gaussian incoming signal with $\tau=1\mu$s would undergo 25$\%$ stretching, with less 0.3$\%$ recovery efficiency reduction. In other words, the stored light would be retrieved within a time interval $\approx$1.6 times shorter than with $\tau=2\mu$s. Hence, with unchanged background level and input pulse energy, the signal-to-background ratio would raise by factor $\approx$1.6. Equivalently, the initial signal-to-background ratio could be recovered with $\approx$1.6 times less photons in the incoming signal.

In comparison with a similar storage scheme, involving two rephasing pulses~\cite{mcauslan2011}, the background level has been strongly reduced. As already noticed, our detector is not exposed to the FID radiation induced by the intense pulses that counterpropagate with the signal. Our results can also be examined in the light of a recent fully developed AFC-based memory demonstration, where storage involves the intermediate conversion of optical coherences into spin coherences, and where readout is triggered by an intense optical pulse~\cite{Timoney2013}. That work gets closer than ours to the single photon operating conditions, with signal to noise ratio one order of magnitude larger than in our experiment. However the very low spurious light background they are able to reach - about 300 times smaller than we do - is somehow balanced by the smallness of the recovery efficiency, 30 times weaker than in our case.    

\section{Conclusion}
Investigating the noise features of the ROSE scheme, we have reached a background level of $\approx$1 photon in the echo mode, evenly shared between spontaneous emission and coherent noise. In these conditions, a 14 photon signal can be recovered with unit signal-to-background ratio. One could probably dispose of the coherent contribution to the noise by anti-reflection coating the crystal. The main limitation of the present work comes from the strong pulse failure to transport the atoms to the upper level and back with the expected efficiency. The rather high level of spontaneous emission in the signal mode results from this feature that should be clarified.         

The research leading to these results has received funding from the People Programme (Marie Curie Actions) of the European Union's Seventh Framework Programme FP7/2007-2013/ under REA grant agreement no. 287252. Support has also been received from the European Union through Project No. FP7-QuReP(STREP-247743), from the Agence Nationale de la Recherche through Project No. ANR-09-BLAN-0333-03 and from the Direction G\'en\'erale de l$'$Armement.

\end{document}